\documentclass[sigplan,screen,nonacm,balance=false]{acmart}
\usepackage{amsmath}
\usepackage{algorithmic}
\usepackage{multirow}
\usepackage{tabularx}
\usepackage{enumitem}
\usepackage{placeins}
\usepackage{float}
\usepackage{subcaption}

\newcommand{\framework}{VortexSynth}
\newcommand{\benchmarks}{VX-Robotics}
\newif\ifshowcomments
\showcommentsfalse

\definecolor{HPColor}{HTML}{2F5597}

\AtBeginDocument{\hypersetup{hidelinks}}

\setcopyright{none}
\acmConference{}{}{}
\acmBooktitle{}
\acmDOI{}
\acmISBN{}

\title[\framework{}]{\framework{}: Profile-guided benchmark reconstruction for GPU architecture research}
\author{Huanzhi Pu}
\authornote{Huanzhi Pu and Chulhyung Park contributed equally to this work.}
\affiliation{%
  \institution{Georgia Institute of Technology}
  \city{Atlanta}
  \state{Georgia}
  \country{USA}}
\email{hpu8@gatech.edu}

\author{Chulhyung Park}
\authornotemark[1]
\affiliation{%
  \institution{Georgia Institute of Technology}
  \city{Atlanta}
  \state{Georgia}
  \country{USA}}
\email{chp@gatech.edu}

\author{Chihyo Ahn}
\affiliation{%
  \institution{Georgia Institute of Technology}
  \city{Atlanta}
  \state{Georgia}
  \country{USA}}
\email{ahnch@gatech.edu}

\author{Ramyad Hadidi}
\affiliation{%
  \institution{d-Matrix}
  \country{USA}}
\email{ramyad.hadidi@gmail.com}

\author{Hyesoon Kim}
\affiliation{%
  \institution{Georgia Institute of Technology}
  \city{Atlanta}
  \state{Georgia}
  \country{USA}}
\email{hyesoon@cc.gatech.edu}

\renewcommand{\shortauthors}{Pu, Park, et al.}

\hypersetup{keeppdfinfo}

\begin{document}
\begin{abstract}
GPU architecture research requires representative benchmarks that
can execute on new designs. However, many vendor-tailored workloads and
library kernels cannot run directly on research architectures because their
implementations are closed-source or rely on programming models unsupported
by the target architecture. Migrating existing kernels requires manual effort
to adapt their implementations and validate their behavior. This creates a
need to automate workload migration to target platforms.

We present \framework{}, a framework that uses profile-guided benchmark
reconstruction to construct GPU benchmarks from reference workloads. Our workflow selects
hot kernels through profiling, synthesizes implementations in the target programming language, verifies
output equivalence, and measures profile differences using a translated proxy implementation.
The resulting profile signature gap is supplied to an agent for iterative refinement. We evaluate
the method on Vortex, an open-ISA GPGPU, using robotics as the application
domain. All active kernel pairs pass both output-equivalence checks, and
the case studies show that iterative refinement can reduce profile distance.
\end{abstract}

\maketitle

\section{Introduction}
\label{sec:intro}

Machine learning, scientific simulation, and other compute-intensive
applications increasingly rely on GPUs for parallel computation and high
memory throughput.
These growing demands motivate new GPU architectures and changes to existing designs.
Several open-source GPU designs have been proposed~\cite{tine2021micro,li2024ventus,balasubramanian2015miaow}.
For example, Vortex is an open-source GPGPU based on RISC-V.
These open GPU designs enable reproducible research and transparency in architectural exploration.
Evaluating architectural ideas and optimizations requires
workloads that exercise the behavior they are intended to improve.
Researchers therefore need representative
benchmarks that can run on their specific architectures.
As applications evolve, new models, operators, and numerical
formats change the kernels these benchmarks must represent. In addition,
new GPU architectures may require existing kernel implementations to be
adapted or retuned~\cite{sereda2026kforge,su2026cudal2}. Benchmark
preparation is therefore a recurring effort as both workloads and target
architectures change.

However, researchers starting work on a new
design have a limited set of benchmarks to use as
baselines, as many vendor-tailored workloads and library kernels do not
run directly on a research architecture. Migrating an existing workload requires adapting its kernels and
library dependencies to the target, then validating the resulting
implementation. Compiler support can reduce part of this work. For example,
VOLT translates CUDA and OpenCL programs for Vortex and its
variants~\cite{jeong2026volt}, but incompatibilities due to hardware
differences (e.g., tensor-core and asynchronous-copy support)
can still require substantial manual adaptation.

An open-source library may still rely on closed-source kernels.
For example, deep-learning frameworks call NVIDIA GPU libraries such as
cuBLAS and cuDNN~\cite{nvidia2026cublas,nvidia2026cudnnfrontend}.
When the required kernel implementation is unavailable, porting the surrounding
code is not enough. Researchers must implement replacement kernels for the
target architecture, adding to the manual effort needed to prepare benchmarks.
Making these computations available as benchmarks on open GPUs would let
researchers evaluate hardware, compiler, and runtime changes using
representative application workloads.

\begin{figure*}[t]
\centering
\includegraphics[width=\textwidth]{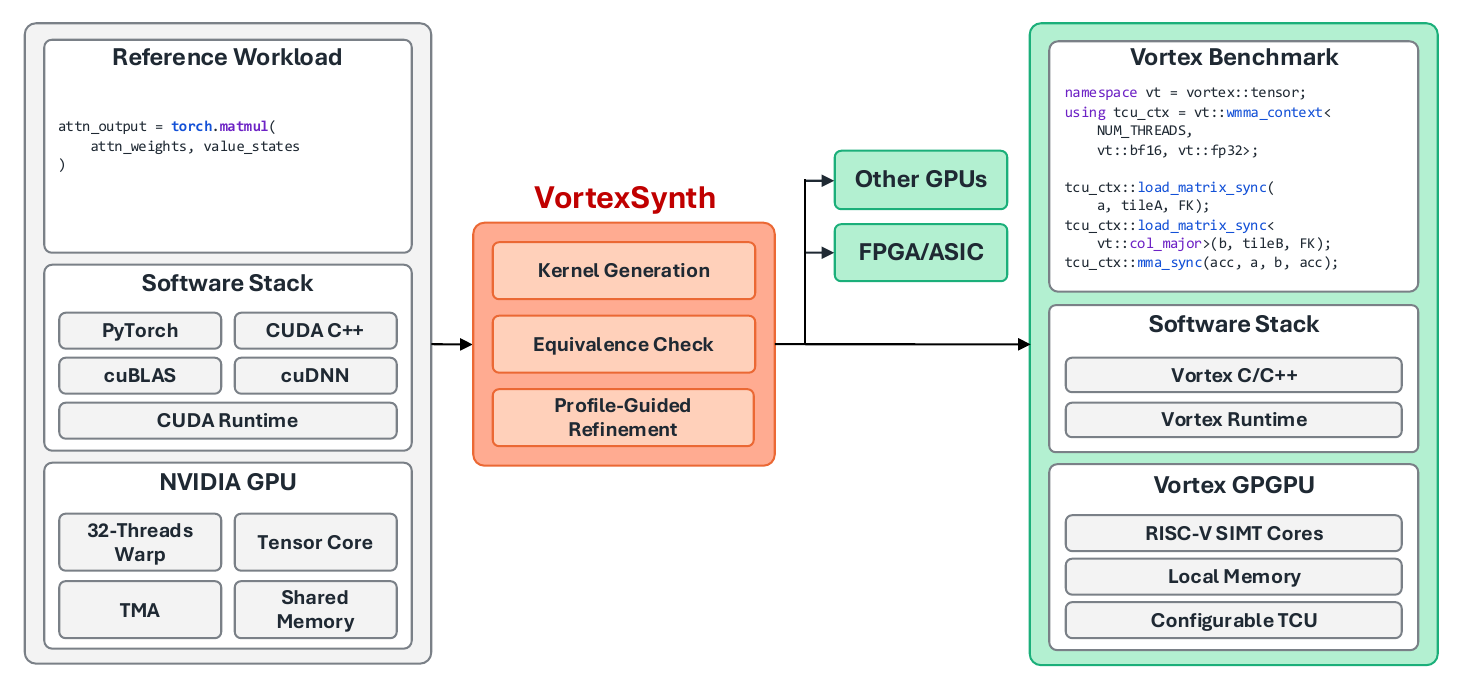}
\caption{From a vendor-dispatched reference kernel to a Vortex
  benchmark using \framework{}.}
\label{fig:framework-overview}
\Description{A CUDA reference workload and its software and hardware stack feed VortexSynth kernel synthesis, output equivalence checks, and profile-based refinement. The generated Vortex benchmark uses the Vortex software and hardware stack. Other GPUs, FPGAs, and ASICs illustrate potential target accelerators.}
\end{figure*}

Figure~\ref{fig:framework-overview} presents an overview of \textbf{\framework{}},
a portable benchmark reconstruction framework that migrates machine-learning
workloads, such as robotics policies, to a target platform.
Our workflow selects hot kernels through profiling, synthesizes implementations
in the target programming language, verifies output equivalence, and measures
profile differences using a translated proxy implementation.
The resulting signature distance is supplied to an agent for iterative refinement.
We select hot kernels rather than whole models because the resulting Vortex
benchmarks are meant to drive architecture design-space exploration in
simulation, where a single full-model inference can take days. .
In this paper, we focus on porting GPU robotics workloads to Vortex, but the
framework also applies to other accelerators.

Our paper makes the following contributions:
\begin{itemize}[leftmargin=1.3em,itemsep=2pt]
\item We present a profile-guided method for constructing executable GPU benchmarks from workload context and measured
profiles under the target architecture's constraints.
\item Building on this framework, we present \textbf{\framework{}}, a code
agent that synthesizes Vortex kernel implementations from GPU workloads
without access to the original source code.
\item We provide the \textbf{\benchmarks{}} benchmark suite of
26 Vortex kernels derived from robotics workloads profiled on datacenter
and edge GPUs. The suite contributes reusable Vortex/CUDA kernel pairs,
reference profiles, and output-validation artifacts for architecture research.
\end{itemize}

\section{Background}
\label{sec:bg}

\subsection{Vortex GPGPU}
\label{sec:bg:vortex}

Vortex is a full-stack open-source RISC-V GPGPU research platform with
SIMT execution, OpenCL software support, simulator and RTL flows, and a
modifiable compiler, driver, and runtime stack~\cite{tine2021micro}.
Its recent release adds a Tensor Core Unit extension with a tensor
API and low-precision multiply-add support~\cite{vortex23release}. We choose Vortex as the target for
profile-guided benchmark reconstruction, as the architecture supports mechanisms including SIMT scheduling, memory
system changes, atomics, and tensor-core extensions. Vortex is the evaluation target in this paper. Applying the method to another GPU would require its programming interface, implementation constraints, and validation tools.

\subsection{Vortex Benchmark}
\label{sec:bg:vortex-runtime}
A Vortex benchmark is a C/C++ benchmark
that targets the Vortex software stack. The host code uses Vortex runtime
calls to allocate device buffers, transfer inputs, launch a kernel, and
copy results back for checking. Compared to a CUDA kernel, the device kernel in a vortex program is compiled by the
Vortex toolchain into RISC-V GPGPU instructions executed by Vortex cores~\cite{vortex23interfaces},
rather than into NVIDIA PTX/SASS or calls to CUDA libraries. Table~\ref{tab:vortex-interfaces} lists the Vortex intrinsics a C/C++ kernel uses for thread launch, lane masking, barriers, local memory, and tensor-core multiply-add.

\begin{table}[t]
\caption{Vortex kernel intrinsics and tensor-core APIs}
\label{tab:vortex-interfaces}
\centering
\scriptsize
\setlength{\tabcolsep}{4pt}
\begin{tabularx}{\columnwidth}{@{}>{\raggedright\arraybackslash}p{0.49\columnwidth} >{\raggedright\arraybackslash}X@{}}
\toprule
\textbf{Vortex intrinsics} & \textbf{Description}\\
\midrule
\texttt{vx\_spawn\_threads(dim,}\newline\texttt{grid, block, kernel, args)} & Launches \texttt{kernel(args)} over a \texttt{dim}-dimensional grid with the given grid and block sizes.\\
\texttt{vx\_thread\_id()}\newline\texttt{vx\_warp\_id()} & Return the current lane and warp identifiers.\\
\texttt{vx\_tmc(mask)} & Sets the active lanes in the current warp using \texttt{mask}.\\
\texttt{vx\_barrier(id, num\_warps)} & Synchronizes \texttt{num\_warps} warps at barrier \texttt{id}.\\
\texttt{\_\_local\_mem(bytes)} & Returns the block's local-memory base using a per-block size of \texttt{bytes}.\\
\texttt{wmma\_context<NT, It, Ot>::}\newline\texttt{mma\_sync(d, a, b, c)} & Computes $d=ab+c$ on the TCU. \texttt{NT} is the warp width, and \texttt{It}/\texttt{Ot} specify input/output types.\\
\bottomrule
\end{tabularx}
\end{table}

\subsection{Robotics Workloads}
\label{sec:bg:robotics-workloads}

We use three widely used robot-learning workloads as the reference applications in this paper: Action Chunking with Transformers (ACT), Diffusion Policy, and $\pi_0$ model.  
ACT predicts a
sequence of future actions from camera observations and the robot's joint
state~\cite{zhao2023act}. Predicting actions together helps capture their
temporal relationships during fine-grained manipulation. Its visual encoder
and transformer policy combine image processing with attention and dense
matrix operations.

Diffusion policies generate actions through repeated refinement
rather than a single prediction. Diffusion Policy starts from a noisy action
sequence and iteratively denoises it while conditioning on visual
observations~\cite{chi2023diffusion}. This formulation can represent multiple
plausible actions for the same observation. Its convolutional implementation
repeatedly applies temporal convolutions, normalization, and activation
functions during action generation.

Vision-language-action (VLA) models combine visual observations and
language instructions to support a range of robot tasks. The $\pi_0$ model
extends a pretrained vision-language model with an action expert that
generates continuous robot actions using flow matching~\cite{rtx2}. It
conditions action generation on the observations, instruction, and robot
state. Its transformer computation involves attention and dense matrix
operations, coupling visual and language processing with action generation.

\subsection{LLM}

LLMs can assist GPU kernel migration by translating source code into
implementations for a different architecture or programming
model~\cite{heakl2026cass,kaplan2026paracodex}. However, translation remains
challenging when the target programming model is less familiar to the model,
and a single generation attempt may not produce a correct
implementation~\cite{davis2025parevalrepo}.
Iterative workflows allow generated implementations to be checked and revised
using compiler diagnostics and execution results. This feedback helps the
model identify and correct errors that remain after the initial
translation~\cite{bitan2025unipar}. These findings motivate the iterative
process in our framework. In addition to output-correctness feedback, we
supply measured profile differences to guide revisions toward closer
agreement with the reference kernel.

\section{\framework{}:Profile-guided benchmark reconstruction}
\label{sec:framework}

\begin{figure*}[t]
  \centering
  \includegraphics[width=\textwidth]{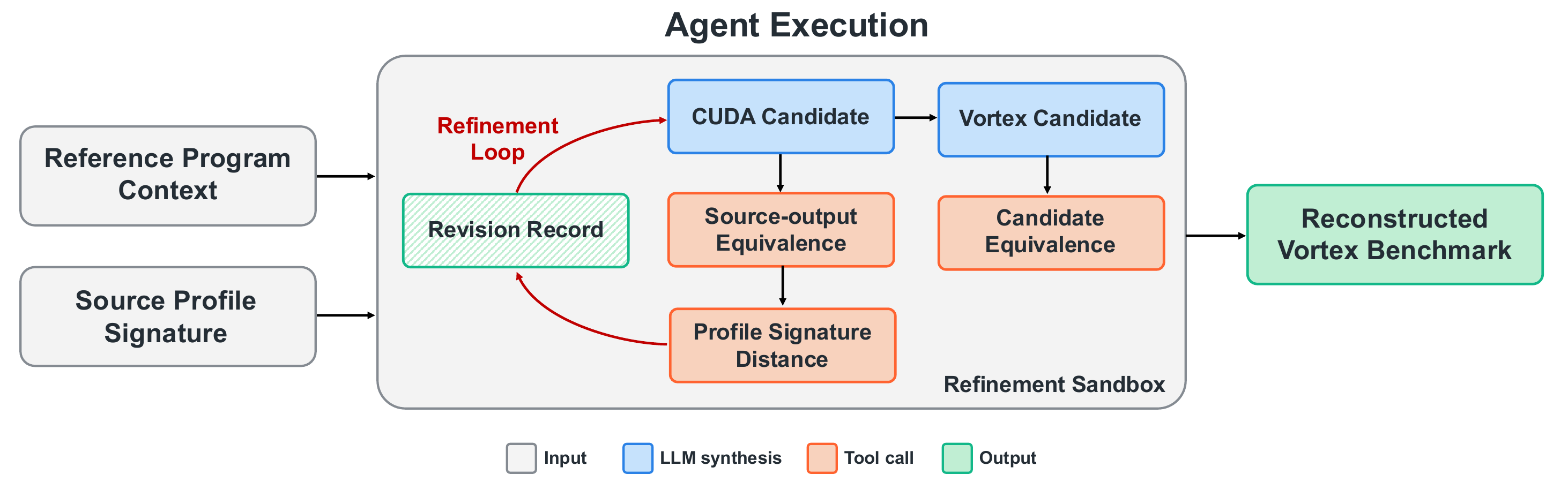}
  \caption{\framework{}: agent structure and refinement loop.}
  \label{fig:workflow}
\Description{The workflow profiles a reference workload, synthesizes Vortex and CUDA candidates, checks output equivalence, and feeds profile differences back into refinement.}
\end{figure*}

Figure~\ref{fig:workflow} illustrates \framework{}, the benchmark reconstruction
workflow. A \emph{revision} is one pass through the refinement loop, in
which an LLM applies an optimization change, generates code, and
records the change. Each revision produces two \emph{candidates}, a
CUDA implementation that runs on an NVIDIA GPU and a Vortex
implementation that runs on SimX, the Vortex simulator. The loop retains a candidate only
when it lowers the profile distance; otherwise it reverts to the last
retained pair. When all optimization axes are exhausted, the loop ends
and the retained candidates are \emph{locked}. The framework delivers
the locked Vortex implementation as the \emph{reconstructed benchmark}.

\subsection{Reference Program Context}
\label{sec:framework:context}




The workflow begins with a reference workload, the target application whose kernels we reconstruct. We first profile the workload at the PyTorch level to record each operator call and the cuBLAS or cuDNN kernel it dispatches to. From the call and its
tensors we take the shapes and the operand and accumulator data types.
From the demangled kernel name we take the engine class, scalar SIMT or
tensor core. From the profiler we take the block size and launch grid.
A property read from the kernel name is used only when the measured
launch geometry agrees with it; properties encoded by undocumented
naming conventions are left for the search to determine.

\subsection{Source Profile Signature}
\label{sec:framework:profile-signature}

Profiling the reference workload produces its \emph{profile signature},
the set of Nsight Systems and Nsight Compute measurements that describe
the reference behavior. It records hot-kernel rank, cumulative time,
launch parameters, and the ten metrics listed in
Table~\ref{tab:similarity-thresholds}. We reconstruct each hot kernel
separately at its profiled shape, which keeps every benchmark within a
simulation budget suited to architecture studies while preserving the
shape that determined the reference's behavior.

\textbf{Profile signature metrics.} The ten Nsight Compute metrics
describe hardware utilization and serve as the behavioral target for
refinement. A benchmark is used to evaluate an architecture or a
compiler change, so what it must preserve is where the kernel puts
pressure, not the exact instruction sequence that produces it. We
therefore group the metrics by the hardware resource an experiment
would vary. Each group describes the pressure that such an experiment would examine. The ten metrics are selected form four
groups. The two cache hit rates and coalescing efficiency describe \emph{where the traffic is served}. Achieved peak compute, tensor MFU, and atomics per instruction describe \emph{which engine does the work and what it costs to commit a result}. Arithmetic intensity and achieved memory bandwidth describe \emph{work per byte and the rate of data transfer}. Active-lane fraction and occupancy describe \emph{how much of the machine the kernel keeps busy}.

Datacenter GPUs such as H200 report all ten metrics, while edge devices such as Jetson AGX Orin omit arithmetic intensity and achieved DRAM bandwidth. Jetson AGX Orin uses a Tegra SoC whose integrated GPU shares
LPDDR5 memory with the CPU through the External Memory Controller (EMC). The integrated GPU does not expose the per-kernel DRAM counters needed for these metrics through the NCU/CUPTI counter API.

\textbf{Signature distance.} We use the signature distance to compare a generated candidate with the reference. Each measured metric contributes a distance between zero and one, and the signature distance $D$ is their mean. Percent metrics use the absolute difference over their range, the two
rate metrics use a log ratio so that a factor of two counts the same in
either direction, and the active-lane fraction and the atomics rate use
the direct difference. A metric the platform cannot measure is left out
of the mean rather than scored as a full distance.

\begin{table}[t]
\caption{Profile-signature metrics by pressure group and their distance.
$c$ is the candidate value and $r$ the reference value; each
$d_m=\min(1,\cdot)$ of the expression shown.
$\varepsilon=10^{-12}$, $C=5$.}
\label{tab:similarity-thresholds}
\centering
\scriptsize
\setlength{\tabcolsep}{4pt}
\begin{tabular}{@{}l l l l@{}}
\toprule
\textbf{Group} & \textbf{Metric} & \textbf{Unit} & \textbf{Distance} $d_m$ \\
\midrule
\multirow{3}{*}{\shortstack[l]{Where traffic\\is served}}
  & L1 hit rate             & \%        & $|c-r|/100$ \\
  & L2 hit rate             & \%        & $|c-r|/100$ \\
  & Coalescing efficiency   & \%        & $|c-r|/100$ \\
\midrule
\multirow{3}{*}{\shortstack[l]{Engine and\\commit}}
  & Achieved peak compute   & \%        & $|c-r|/100$ \\
  & Tensor MFU              & \%        & $|c-r|/100$ \\
  & Atomics per instruction & per inst. & $|c-r|$ \\
\midrule
\multirow{2}{*}{\shortstack[l]{Work per\\byte}}
  & Arithmetic intensity    & FLOP/byte & $|\log_2\frac{c+\varepsilon}{r+\varepsilon}|/C$ \\
  & Achieved memory bw.     & GB/s      & $|\log_2\frac{c+\varepsilon}{r+\varepsilon}|/C$ \\
\midrule
\multirow{2}{*}{\shortstack[l]{Machine\\busy}}
  & Active-lane fraction    & $[0,1]$   & $|c-r|$ \\
  & Occupancy               & \%        & $|c-r|/100$ \\
\midrule
\addlinespace[2pt]
\midrule
\multicolumn{3}{@{}l}{Signature distance $D$} & mean of populated $d_m$ \\
\bottomrule
\end{tabular}
\end{table}


\textbf{Signature scope.} The signature describes which hardware resources
a kernel uses and how heavily it uses them. Similar resource-use patterns
can arise from different instruction schedules, stall reasons,
shared-memory bank conflicts, and register allocations.

Coalescing efficiency is averaged over ordinary global load instructions.
For kernels that move operands through asynchronous or bulk copies, it
reflects only the remaining ordinary loads. If the reference issues none,
we report coalescing efficiency as 0 and retain it in $D$.
Tensor MFU measures tensor-core use.
A candidate that executes entirely on scalar SIMT units has a measured
tensor MFU of zero. We retain this zero in $D$ to capture the difference
in engine use between that candidate and a tensor-core reference.

\subsection{Refinement Sandbox}
\label{sec:framework:sandbox}

An LLM generates the kernel code and then revises it inside a
sandbox defined by a constraint set and a validation procedure. The constraints specify which changes a revision may make, which constructs it may use, and which properties must remain fixed.
The agent may only edit the files named in the task, and it may not
substitute a different algorithm, change the swap target, call the vendor library instead of the generated kernel, replay cached outputs, or tune constants in ways that break correctness.
The validation procedure specifies how a candidate replaces the vendor kernel, how output equivalence is checked, and which tolerance applies to each precision path. These rules are fixed before refinement begins and apply to every revision.

\textbf{Allowed Code optimization axes.} The constraint set defines the agent's search space through the optimization axes in Table~\ref{tab:porting-axes}. Each permitted axis must be expressible on Vortex so that the CUDA and Vortex candidates implement corresponding changes. To make all four profile-pressure groups in Section~\ref{sec:framework:profile-signature} actionable, the search space includes portable optimizations that primarily target each group. Shared-memory staging, read-once staging, and lane-contiguous mapping alter where operands are served and how lanes access them. Tile dimensions, register tiling, and software prefetching adjust data reuse and load--compute overlap. Tensor-core lowering, fused-activation stripping, and split-K alter the execution engine or result-commit behavior, while per-block decomposition, boundary handling, and the generic K range affect exposed parallelism and machine utilization. These twelve axes form a compact, portable search space; Section~\ref{sec:evaluation:profile} measures their actual effects on the profile signature. In particular, four axes express NVIDIA-oriented optimizations using mechanisms available on Vortex. Read-once staging corresponds to cache-residency policy, lane-contiguous mapping replaces vectorized loads, software prefetching replaces asynchronous copies, and split-K without atomics replaces floating-point atomics on the output. Each targets the same metric with ordinary loads and stores. The Vortex tensor unit provides FP16 and BF16 operand formats and no TF32, so a tensor-core axis on a TF32 reference is expressed at FP16.

\begin{table}[t]
\caption{Code Optimization axes admitted during LLM revision. Each axis is expressible on both CUDA and Vortex Benchmarks.}
\label{tab:porting-axes}
\centering
\small
\setlength{\tabcolsep}{3pt}
\renewcommand{\arraystretch}{1.12}
\begin{tabularx}{\linewidth}{@{}r X @{\hspace{1em}} r X@{}}
\toprule
\multicolumn{4}{c}{\textbf{Optimization axes}}\\
\midrule
1 & Per-block decomposition & 7 & Register tiling / ILP \\
2 & Tensor-core lowering & 8 & Software prefetching \\
3 & Tile dimensions & 9 & Boundary handling \\
4 & Shared-memory staging & 10 & Generic K range \\
5 & Read-once staging & 11 & Fused activation stripping \\
6 & Lane-contiguous mapping & 12 & Split-K without atomics \\
\bottomrule
\end{tabularx}
\end{table}

\subsection{Candidate Generation}

\label{sec:framework:synthesis}


Each selected hot kernel is reconstructed independently, with its own
candidates and its own revision chain. The LLM generates code from an
input bundle containing the reference program context, source profile
signature, sandbox definitions, and revision history. All reported runs
used \texttt{Claude Opus 5}~\cite{opus5}.


\textbf{CUDA candidate.} For every revision, the LLM generates a
\emph{CUDA candidate}. The first candidate is built from the reference
program context alone. From that context we extract four reference
properties (Table~\ref{tab:golden-answers}). They remain fixed during
refinement and constrain four of the twelve optimization axes. For
example, per-block decomposition and tile dimensions must preserve the
measured block size. The engine class and the data types bound
tensor-core lowering, and the profiled shape bounds the split factor for
split-K. From the second candidate onward, the LLM enters the refinement
loop and normally applies one axis from Table~\ref{tab:porting-axes} for each
revision.

\begin{table}[t]
\caption{Properties fixed before refinement. } 
\label{tab:golden-answers}
\centering
\footnotesize
\setlength{\tabcolsep}{4pt}
\renewcommand{\arraystretch}{1.1}
\begin{tabularx}{\linewidth}{@{}l l X X@{}}
\toprule
& \textbf{Fixed property} & \textbf{Established from} & \textbf{Bounds}\\
\midrule
1 & Engine class    & kernel name, scalar versus tensor-core tokens
  & Tensor-core lowering\\
2 & Data types      & captured tensors and the name's dtype tokens
  & Tensor-core lowering\\
3 & Block size      & measured launch geometry
  & Per-block decomposition / Tile dimensions\\
4 & Profiled shape  & baseline grid and tile size, cross-checked
  & Split-K without atomics\\
\bottomrule
\end{tabularx}
\end{table}

\textbf{Vortex candidate.} The LLM generates a \emph{Vortex candidate}
based on the CUDA candidate of the same revision under the same input bundle and optimization axis. The constraint set specifies the Vortex construct for each axis, keeping the pair aligned throughout refinement. The CUDA candidate supplies Nsight Compute measurements for the implementation strategy shared by the pair.

\textbf{Swapped execution.} We insert the CUDA candidate into the reference workload by wrapping selected PyTorch operators, such as \nolinkurl{torch.matmul} and \nolinkurl{F.linear}. The wrapper saves the original function and replaces its Python entry point. On each call, it checks the input shapes against the profiled matrix dimensions $(M,N,K)$ or convolution shape. A matching call is routed to \nolinkurl{proxy.forward}, which launches the generated CUDA kernel. Other calls use the saved original function, preserving the rest of the workload. The wrapper logs the call signature and dispatch decision so that the output checker can verify that the candidate actually ran. This lets us validate and profile the replacement within the application.

\subsection{Output Equivalence}
\label{sec:framework:equivalence}

Each revision must pass the source-output equivalence and candidate equivalence checks before we measure its profile signature distance. A precision path comprises the data types used by the reference kernel, the CUDA candidate, and the Vortex candidate. Table~\ref{tab:equivalence-chain} specifies the per-element absolute and relative tolerances for each path.

\textbf{Source-output equivalence.} We compare the CUDA candidate with the original framework or library operation on the same sampled inputs and verify that the replacement actually executed. When an operator-level swap is unavailable, a CPU/PyTorch implementation of the same operation supplies the reference outputs.

\textbf{Candidate equivalence.} We run the CUDA candidate on an NVIDIA GPU and the Vortex candidate on SimX using the same input value, then compare their outputs.

The tolerances reflect the precision path and accumulation differences in each check. FP32 comparisons use tight tolerances, while BF16 comparisons allow for rounding on the BF16 output grid. Tensor-core reference paths use wider relative tolerances to accommodate cuDNN's TF32 execution for some shapes. The CUDA-to-Vortex check also accounts for the change from TF32 to FP16 on Vortex. Section~\ref{sec:evaluation:output-equivalence} reports the observed errors.

\begin{table*}[t]
\caption{Source-output and candidate equivalence tolerance per precision
path, written absolute\,/\,relative and applied per element as
$|d| \le \mathrm{atol} + \mathrm{rtol}\,|y|$.}
\label{tab:equivalence-chain}
\centering
\footnotesize
\setlength{\tabcolsep}{5pt}
\begin{tabular}{@{}l c l c l r@{}}
\toprule
\textbf{Reference kernel} & \textbf{Source-output eq.} & \textbf{CUDA candidate} &
\textbf{Candidate eq.} & \textbf{Vortex candidate} & \textbf{Pairs}\\
\midrule
FP32 & $\xrightarrow{10^{-3}\,/\,10^{-3}}$ & FP32 &
  $\xrightarrow{10^{-3}\,/\,10^{-3}}$ & FP32 & 9\\
TF32 & $\xrightarrow{10^{-3}\,/\,2.5{\times}10^{-1}}$ & TF32 &
  $\xrightarrow{2{\times}10^{-3}\,/\,2.5{\times}10^{-1}}$ & FP16 & 7\\
BF16 & $\xrightarrow{10^{-2}\,/\,10^{-2}}$ & BF16 &
  $\xrightarrow{10^{-2}\,/\,10^{-2}}$ & BF16 & 10\\
\bottomrule
\end{tabular}

\end{table*}

\subsection{Profile-Guided Refinement}
\label{sec:framework:verification-validation}

Each refinement revision starts from the best retained design. The LLM normally selects one optimization axis from Table~\ref{tab:porting-axes}, applies it to the retained design, and generates the corresponding CUDA and Vortex candidates. When single-axis attempts no longer improve the profile, the agent may combine two permitted axes in a revision. The harness then runs the source-output equivalence and candidate equivalence checks. A failed check rejects the revision. Otherwise, the harness profiles the CUDA candidate within the workload and computes its signature distance $D$.

A revision is \emph{retained} if it lowers $D$ by more than the measurement-noise margin, $\delta=0.005$. Otherwise, it is \emph{reverted}. For revisions that apply a single axis, the change in $D$ from the retained design is attributable to that axis. The per-metric changes show which parts of the signature the axis affects, forming the basis of the analysis in Section~\ref{sec:evaluation:profile}. The next revision starts from the retained design and receives the per-metric distances, so the agent can choose an axis that addresses the largest remaining differences.

The agent decides when to stop based on the revision history and remaining profile differences. It typically stops after trying the applicable axes and judging that no further change is likely to improve profile agreement. The agent can revisit an axis through a different implementation or direction of change, so one failed attempt does not exhaust that axis. The best retained pair is then finalized as the benchmark and its CUDA proxy. We report the final distance and revision history to show the agreement reached under these constraints.

\textbf{Revision record.} Each kernel has a revision record that is supplied to its next revision. The record identifies the applied optimization and the output or profile condition it targets. It also documents why the candidate still implements the same kernel, the outcome of both output checks, changes in the profile and signature distance, and the retain or revert decision. Reverted attempts remain in the record so the agent can see which changes did not help and avoid repeating them. This record connects each applied axis to its measured effect and informs the next refinement step.

\section{Evaluation}
\label{sec:evaluation}


We use robotics as the evaluation domain, testing whether profiles from
ACT, Diffusion Policy, and Pi0 can guide the construction of Vortex
benchmarks. All workloads are profiled at batch size one, matching the
online inference regime. The evaluation proceeds in five steps. First,
we measure how much H200 and Jetson AGX Orin GPU time the selected hot
kernels cover, and list the suite they yield. Second, we confirm that
every benchmark passes both output checks. Third, we read the revision
records as experiments, measuring which pressure group of
Section~\ref{sec:framework:profile-signature} each optimization axis
moves, and in which direction, for each benchmark type. Fourth, we
follow one benchmark through its full revision chain as a case study.
Fifth, we reverse the order of refinement and iterate on the Vortex
candidate first, to test whether refining on the reference GPU is what
makes the profile signal usable, and read that result against cross-GPU
metric variation.

\subsection{Kernel Coverage}
\label{sec:evaluation:coverage}

\begin{figure}[t]
\centering
\includegraphics[width=\columnwidth]{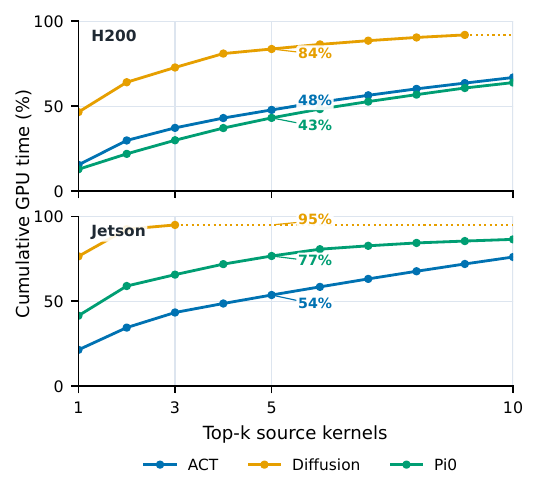}
\caption{Hot-kernel coverage for selection. The vertical line marks the top-five hot-kernel
cutoff. Dotted line segments extend workloads with fewer than ten
listed hot kernels.}
\label{fig:case-hot-kernel-coverage}
\Description{Cumulative GPU-time coverage by hot-kernel rank for each workload and platform. A vertical line marks the top-five cutoff.}
\end{figure}

We select the top five kernels by GPU time from each workload on each platform. Figure~\ref{fig:case-hot-kernel-coverage} shows their cumulative share of GPU time. The resulting suite contains 26 benchmarks covering linear layers, matrix multiplication, and 1-D and 2-D convolution. Table~\ref{tab:kernel-inventory} lists each benchmark's reference kernel, GPU-time share, precision path, revision count, and final signature distance.

\begin{table*}[t]
\caption{The 26 benchmarks of the suite, grouped by type, with the
outcome of refinement for each.}
\label{tab:kernel-inventory}
\centering\scriptsize\setlength{\tabcolsep}{2.4pt}
\begin{tabular}{@{}l l l l l l r c c c r l r@{}}
\toprule
& & & & & & \textbf{GPU} & \multicolumn{3}{c}{\textbf{Precision path}} & \textbf{\# of} & \textbf{LOCKED} & \\
\cmidrule(lr){8-10}
\textbf{Type} & \textbf{Benchmark} & \textbf{Workload} & \textbf{Platform} &
\textbf{Reference kernel} & \textbf{Library} & \textbf{time$^\dagger$ } &
\textbf{Ref.} & \textbf{CUDA} & \textbf{Vortex} & \textbf{revs.} & \textbf{rev.} & \textbf{$D$}\\
\midrule
Linear & \texttt{Linear-1} & ACT & H200 & \texttt{sm80\_xmma\_gemm\_f32 32x32x8 s3} & cuBLAS & 15.4\,\% & FP32 & FP32 & FP32 & 13 & \texttt{r11} & 0.0374\\
 & \texttt{Linear-2} & ACT & H200 & \texttt{cutlass\_80\_simt\_sgemm 128x64 tn} & CUTLASS & 14.4\,\% & FP32 & FP32 & FP32 & 10 & \texttt{r4} & 0.0798\\
 & \texttt{Linear-3} & ACT & H200 & \texttt{sgemm\_largek\_lds64} & cuBLAS & 7.4\,\% & FP32 & FP32 & FP32 & 7 & \texttt{r1} & 0.0533\\
 & \texttt{Linear-4} & ACT & H200 & \texttt{sm80\_xmma\_gemm\_f32 64x64x8 s3} & cuBLAS & 5.8\,\% & FP32 & FP32 & FP32 & 10 & \texttt{r4} & 0.0574\\
 & \texttt{Linear-5} & ACT & Orin & \texttt{ampere\_sgemm 128x64 tn} & cuBLAS & 21.4\,\% & FP32 & FP32 & FP32 & 8 & \texttt{r6} & 0.0689\\
 & \texttt{Linear-6} & ACT & Orin & \texttt{ampere\_sgemm 128x32 tn} & cuBLAS & 13.1\,\% & FP32 & FP32 & FP32 & 9 & \texttt{r8} & 0.0508\\
 & \texttt{Linear-7} & Pi0 & H200 & \texttt{nvjet\_tst 128x272 coopA} & cuBLAS-Lt & 9.0\,\% & BF16 & BF16 & BF16 & 6 & \texttt{r0} & 0.2071\\
 & \texttt{Linear-8} & Pi0 & H200 & \texttt{nvjet\_tst 64x32 h} & cuBLAS-Lt & 7.2\,\% & BF16 & BF16 & BF16 & 9 & \texttt{r6} & 0.1461\\
 & \texttt{Linear-9} & Pi0 & H200 & \texttt{nvjet\_tst 40x64 v} & cuBLAS-Lt & 5.9\,\% & BF16 & BF16 & BF16 & 7 & \texttt{r3} & 0.1539\\
 & \texttt{Linear-10} & Pi0 & Orin & \texttt{ampere\_bf16\_s16816gemm 128x128 s32x5} & cuBLAS & 41.5\,\% & BF16 & BF16 & BF16 & 5 & \texttt{r1} & 0.1952\\
 & \texttt{Linear-11} & Pi0 & Orin & \texttt{ampere\_bf16\_s16816gemm 128x64 s64x3} & cuBLAS & 17.5\,\% & BF16 & BF16 & BF16 & 4 & \texttt{r1} & 0.1731\\
 & \texttt{Linear-12} & Pi0 & Orin & \texttt{ampere\_bf16\_s16816gemm 64x64 sliced1x2 s64x5} & cuBLAS & 4.8\,\% & BF16 & BF16 & BF16 & 6 & \texttt{r3} & 0.0648\\
\midrule
Matmul & \texttt{Matmul-1} & Pi0 & H200 & \texttt{cutlass\_75\_wmma\_tensorop\_bf16 32x32 nn} & CUTLASS & 12.7\,\% & BF16 & BF16 & BF16 & 9 & \texttt{r8} & 0.0820\\
 & \texttt{Matmul-2} & Pi0 & H200 & \texttt{sm90\_xmma\_gemm\_bf16 64x64x64} & cuBLAS & 8.0\,\% & BF16 & BF16 & BF16 & 7 & \texttt{r6} & 0.0501\\
 & \texttt{Matmul-3} & Pi0 & Orin & \texttt{cutlass\_75\_wmma\_tensorop\_bf16 32x32 tn} & CUTLASS & 6.7\,\% & BF16 & BF16 & BF16 & 7 & \texttt{r1} & 0.0698\\
 & \texttt{Matmul-4} & Pi0 & Orin & \texttt{cutlass\_75\_wmma\_tensorop\_bf16 32x32 nn} & CUTLASS & 6.2\,\% & BF16 & BF16 & BF16 & 7 & \texttt{r2} & 0.0537\\
\midrule
Conv2d & \texttt{Conv2d-1} & ACT & H200 & \texttt{sm80\_xmma\_fprop\_tf32 64x32x64 s5} & cuDNN & 4.8\,\% & TF32 & TF32 & FP16 & 6 & \texttt{r0} & 0.2489\\
 & \texttt{Conv2d-2} & ACT & Orin & \texttt{sm86\_xmma\_fprop\_tf32 indexed 128x64x32 s4} & cuDNN & 5.3\,\% & TF32 & TF32 & FP16 & 6 & \texttt{r4} & 0.2008\\
 & \texttt{Conv2d-3} & ACT & Orin & \texttt{cudnn\_ampere\_scudnn 128x64 relu} & cuDNN & 5.0\,\% & FP32 & FP32 & FP32 & 11 & \texttt{r10} & 0.0376\\
 & \texttt{Conv2d-4} & Diffusion & H200 & \texttt{sm80\_xmma\_fprop\_tf32 64x32x64 s5} & cuDNN & 8.2\,\% & TF32 & TF32 & FP16 & 6 & \texttt{r3} & 0.1831\\
\midrule
Conv1d & \texttt{Conv1d-1} & Diffusion & H200 & \texttt{sm90\_xmma\_fprop\_tf32 64x64x32} & cuDNN & 8.7\,\% & TF32 & TF32 & FP16 & 5 & \texttt{r1} & 0.1555\\
 & \texttt{Conv1d-2}$^\ddagger$ & Diffusion & H200 & \texttt{cudnn dgrad\_engine} & cuDNN & 2.7\,\% & FP32 & FP32 & FP32 & 9 & \texttt{r3} & 0.1100\\
 & \texttt{Conv1d-3} & Diffusion & Orin & \texttt{sm86\_xmma\_fprop\_tf32 128x128x16 s3} & cuDNN & 16.0\,\% & TF32 & TF32 & FP16 & 9 & \texttt{r7} & 0.1891\\
 & \texttt{Conv1d-4} & Diffusion & Orin & \texttt{sm80\_xmma\_fprop\_tf32 indexed wo\_smem 128x32x64} & cuDNN & 2.5\,\% & TF32 & TF32 & FP16 & 10 & \texttt{r3} & 0.0975\\
 & \texttt{Conv1d-5} & Diffusion & Orin & \texttt{sm86\_xmma\_fprop\_tf32 indexed 128x64x32 s4} & cuDNN & 1.0\,\% & TF32 & TF32 & FP16 & 7 & \texttt{r4} & 0.2504\\
 & \texttt{Conv1d-6}$^\ddagger$ & Diffusion & Orin & \texttt{cudnn dgrad\_engine} & cuDNN & 0.6\,\% & FP32 & FP32 & FP32 & 7 & \texttt{r5} & 0.1019\\
\bottomrule
\end{tabular}

\vspace{2pt}\footnotesize
Reference kernel names are shortened to the family and tile or stage
signature, and alignment and epilogue suffixes are omitted. $^\dagger$Share of GPU time within each workload on the given platform.
$^\ddagger$Transposed convolution (\texttt{ConvTranspose1d}), grouped with
Conv1d.
\end{table*}

\subsection{Output Equivalence Results}
\label{sec:evaluation:output-equivalence}

All 26 benchmarks pass both output checks under the tolerances in Table~\ref{tab:equivalence-chain}. Against the reference operation, the largest errors are $1.6{\times}10^{-6}$ absolute for FP32, $1.1{\times}10^{-2}$ relative for TF32, and $4.9{\times}10^{-4}$ absolute for BF16.

Across CUDA and Vortex, all nine FP32 pairs pass the candidate equivalence check within the specified tolerance. The TF32-to-FP16 pairs have the largest relative error, $9.5{\times}10^{-2}$, reflecting the change in precision. BF16 pairs reach $3.2{\times}10^{-2}$ relative error on near-zero outputs and pass the combined absolute and relative tolerance. Near zero, the absolute term allows small rounding differences that appear large when expressed as relative error. These checks establish output agreement for the tested inputs across all three precision paths.

\subsection{How Optimization Axes Move the Signature}
\label{sec:evaluation:profile}

We use revisions that apply one optimization axis to examine how each
axis changes the profile signature. For each revision, we compare its
metrics with those of its retained parent using the distance transform
in Table~\ref{tab:similarity-thresholds}, keeping the sign of the change.
Percentage metrics are expressed as fractions of their range, and rate
metrics use a log ratio. Both retained and reverted revisions contribute
to this analysis because an unsuccessful change still shows how an axis
affects the measured behavior.

\begin{figure*}[t]
\centering
\includegraphics[width=\textwidth]{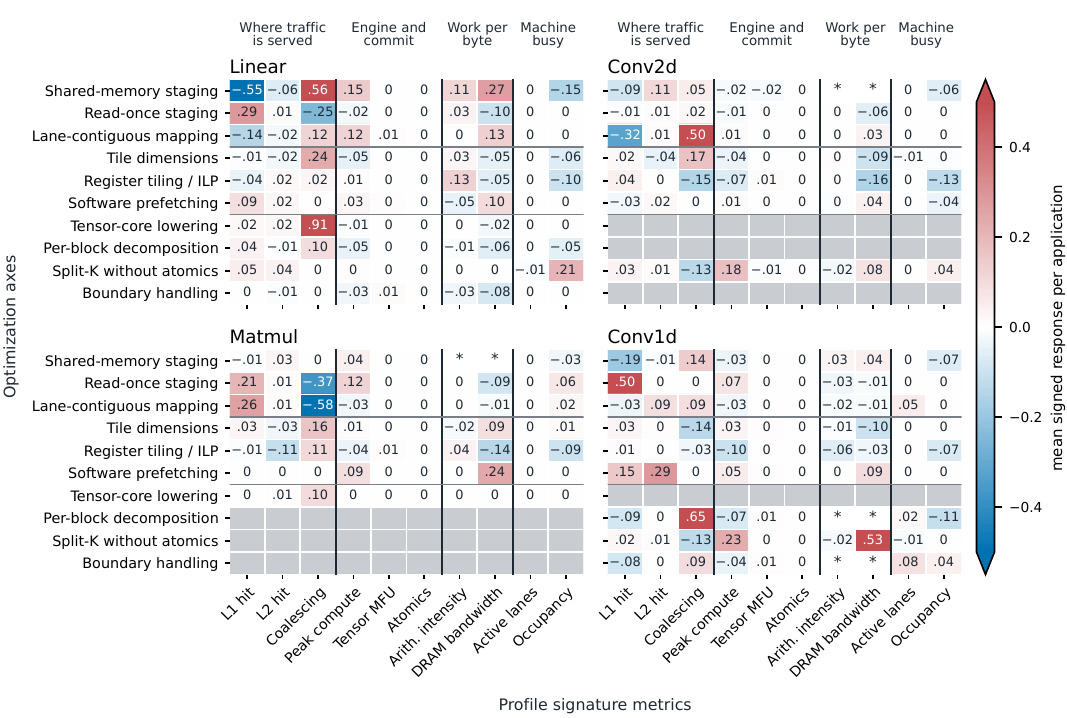}
\caption{Mean signed metric response to each optimization axis, grouped by benchmark type and pressure group. Rows follow the order of the discussion, with rules between the three groups of axes.}
\label{fig:axis-response}
\Description{Four heatmaps, one per benchmark type, with the ten
optimization axes the loop applied as rows and the ten signature
metrics as columns, grouped by the behavior each puts pressure on.
Staging, read-once staging and lane mapping move the cache and
coalescing columns, register tiling, tile dimensions and prefetching
move arithmetic intensity, bandwidth and occupancy, the atomics column
is zero everywhere, and the tensor-utilization and active-lane columns
stay near zero. Gray rows mark axes the loop never applied alone on a
type.}
\end{figure*}

Figure~\ref{fig:axis-response} summarizes these responses by benchmark
type. Rows represent optimization axes, and columns represent the ten
metrics organized into the four pressure groups of
Section~\ref{sec:framework:profile-signature}. Each cell is the mean
signed response across revisions that applied that axis alone. Red
indicates an increase and blue a decrease. Either direction can improve
agreement, depending on the reference value. A value near zero can
reflect small changes or opposing responses across revisions.

The figure includes 157 revisions across 26 benchmarks, with 90 from
H200 and 67 from Orin. We exclude 13 revisions that combined two axes
because their responses cannot be attributed to one axis. We also exclude
three revisions that restored a fixed property and one re-measurement. Generic
K range and fused activation stripping are determined by the profiled
shape and operator in the seed, so neither appears as a refinement row.
Gray rows indicate that an axis was never applied alone for that
benchmark type. Arithmetic intensity and DRAM bandwidth use H200
measurements only. Asterisks mark cells whose revisions ran only on Orin,
where those metrics were unavailable.

Staging and lane mapping primarily affect \emph{where traffic is
served}. Across these axes, the mean magnitude of the plotted responses
is 0.16 for the traffic metrics, compared with 0.05 for \emph{work per
byte}. The linear-layer panel shows the clearest example. \textbf{Shared-memory
staging} lowers the L1 hit rate by 0.55 and raises coalescing by 0.56 on
average. This pattern is consistent with moving repeated reads into
shared memory and organizing the remaining global loads into contiguous
cooperative fills. \textbf{Read-once staging} moves these metrics in the opposite
direction on linear layers and matmuls, as repeated global requests can
be served by L1.

The reference determines which direction is useful for \textbf{lane-contiguous mapping}. On matmuls, five
of the seven lane-mapping revisions reduce coalescing, and all five are
retained. Applying the axis in reverse helps match references that read
few useful bytes per memory sector. Read-once staging provides another
example. It is retained on three matmul benchmarks but reverted on
\texttt{Linear-2} and \texttt{Linear-4}, whose references have L1 hit
rates near zero. Raising L1 hit rate moves those linear-layer candidates
away from their references.

\textbf{Tile dimensions}, \textbf{register tiling / ILP}, and \textbf{software prefetching} connect
\emph{work per byte} with \emph{where traffic is served}. Their mean
response magnitudes are similar in the two groups, about 0.06 in each.
Register tiling increases arithmetic intensity on linear layers as each
thread reuses inputs across a larger output tile. It also lowers mean
occupancy by 0.07--0.13 across the four benchmark types, consistent with
greater register demand per thread. Tile dimensions is the most
frequently applied axis, but its effects vary across benchmark types.
Changing a tile's reduction depth, area, or aspect ratio changes both
reuse and memory access. Mean coalescing rises on linear layers, matmuls,
and 2-D convolutions, but falls on 1-D convolutions.

Software prefetching illustrates why these effects must be evaluated
together. It raises mean DRAM bandwidth in all four panels, yet all seven
applications to 1-D convolutions are reverted. Six increase the overall
distance, and one lowers it by less than the retention margin. The
1-D convolution panel also shows substantial changes in cache hit rates.
Overlapping loads with computation can therefore improve bandwidth
without improving agreement with the reference's full profile.

The \emph{engine and commit} and \emph{machine busy} groups respond less
strongly overall, although their metrics behave differently. Atomics
remain zero because the references use none and the sandbox excludes
them. Tensor MFU changes little, and active-lane fraction is nearly
unchanged outside the 1-D convolution cases. These patterns are
consistent with preserving the engine class, block size, and shape.
Occupancy is more responsive because several axes change resource use
or the amount of work in flight. Shared-memory staging, for example,
reduces mean occupancy on linear layers by 0.15. Achieved peak compute
also changes when a revision changes execution time for the same amount
of work.

Fixed reference properties still leave room for changes in other
pressure groups. The \textbf{tensor-core lowering} axis changes fragment shape while keeping the engine class fixed. This changes the lane-to-element mapping. In the two
linear-layer applications of this axis, the largest mean response is in
coalescing, at 0.91. \textbf{Per-block decomposition} similarly changes the
arrangement of warps over a tile while preserving block size, with its
largest mean response also in coalescing on 1-D convolutions. \textbf{Split-K
without atomics} instead raises DRAM bandwidth and achieved peak compute
on 1-D convolutions by exposing parallelism along the reduction
dimension. The measured effect of an axis can therefore extend beyond
the pressure group it was intended to target.

\subsection{Case Study: Pi0 Attention Matmul}
\label{sec:evaluation:refinement}

\begin{figure}[t]
\centering
\includegraphics[width=\columnwidth]{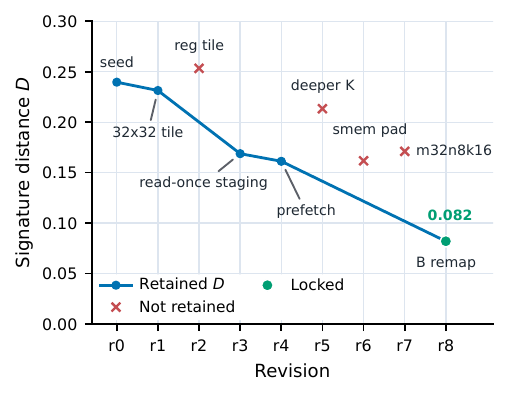}
\caption{Signature distance $D$ across CUDA-candidate revisions of the Pi0
BF16 tensor-core matmul (\texttt{Matmul-1}) on NVIDIA H200.}
\label{fig:signature-distance-trace}
\Description{Signature distance over nine measured Pi0 CUDA-candidate
revisions on H200. The retained trace falls from 0.240 to 0.082, with
crosses marking the four reverted attempts and short labels naming the
change each revision made.}
\end{figure}

We follow the top kernel of the Pi0 stack on H200, the BF16
tensor-core matmul that weights values by attention scores
(\texttt{Matmul-1}), and evaluate
each stage of its refinement against the map of
Section~\ref{sec:evaluation:profile}.
Table~\ref{tab:h200-pi0-k1-deep-dive} and
Figure~\ref{fig:signature-distance-trace} give the chain. Every
revision preserves output equivalence. Its measured response can be compared with the mean response of the corresponding axis in the Matmul panel of Figure~\ref{fig:axis-response}.

\textbf{Seed (r0).} The seed preserves the four reference properties listed in Table~\ref{tab:golden-answers}. It uses a tensor-core engine, BF16 operands with FP32 accumulation, 128 threads per block, and the profiled batched matrix shape. These properties constrain the engine and machine-busy groups before refinement.
Occupancy, active lanes, tensor MFU and atomics sit within 0.02 of the
reference at r0 and no revision moves them, which is the inheritance
Section~\ref{sec:evaluation:profile} describes. What the seed misses is
where the traffic is served, with an L1 hit rate of 14\% against 89\%,
and the work per byte, with an arithmetic intensity near a quarter of
the reference's.

\textbf{Closing the traffic group (r1--r4).} The three retained
revisions improve cache behavior and recover bandwidth, consistent with the responses in the Matmul
panel. Stating the $32\times32$ tile (r1) and read-once staging
(r3) raise the L1 hit rate, the second by 0.52 in one step, and
software prefetching (r4) recovers the DRAM bandwidth that staging
cost. The one reverted revision, register tiling (r2), did not move
arithmetic intensity under a doubling of reuse, so the additional reuse did not reduce measured DRAM traffic in this revision.
Read-once staging is also where the reference sets the sign of an
axis. The same construct was reverted on \texttt{Linear-2} and
\texttt{Linear-4}, whose references sit near zero L1 hit rate.

\textbf{Unsuccessful refinement attempts (r5--r7).} Deepening the staged reduction,
padding the staged tiles, and changing the tensor-core fragment shape
each raise $D$ or fail to lower it by the noise margin. The third is an
admitted axis applied against a fixed property, since only the
fragment shape inside the fixed engine class moves, and it registers
as coalescing alone. It is the one application behind the tensor-core
row of the Matmul panel. These three changes fail to improve profile
agreement, leaving r4 as the best retained revision.

\textbf{Reversing lane mapping (r8).} The agent revisits lane-contiguous mapping (axis 6 in Table~\ref{tab:porting-axes}), this time applying it in reverse. This choice targets the largest remaining profile difference. The retained r4 candidate is more coalesced than the reference, so reducing coalescing can bring its profile closer to the reference. The agent remaps the second operand's loads so that consecutive lanes access separate memory sectors while preserving the staged values and their order. Coalescing falls to 29\%, closer to the reference's 37\%, and the signature distance falls to $D=0.0820$. The revision is retained because this change improves profile agreement while preserving output equivalence.

\begin{table}[!t]
\caption{CUDA-candidate refinement of Pi0 attention matmul on H200.}
\label{tab:h200-pi0-k1-deep-dive}
\centering
\footnotesize
\setlength{\tabcolsep}{3pt}
\begin{tabular}{@{}l l c c l@{}}
\toprule
\textbf{Rev.} & \textbf{Axis applied} & \textbf{$D$} & \textbf{$\Delta D$} & \textbf{Action} \\
\midrule
r0    & Seed, no axis applied        & 0.2397 & --      & baseline \\
r1    & Tile dimensions              & 0.2314 & $-$0.0082 & retain \\
r2    & Register tiling / ILP        & 0.2534 & $+$0.0220 & revert \\
r3    & Read-once staging            & 0.1688 & $-$0.0626 & retain \\
r4    & Software prefetching         & 0.1613 & $-$0.0075 & retain \\
r5    & Tile dimensions, $K$ depth   & 0.2134 & $+$0.0521 & revert \\
r6    & Lane-contiguous mapping      & 0.1617 & $+$0.0005 & revert \\
r7    & Tensor-core lowering         & 0.1711 & $+$0.0098 & revert \\
r8    & Lane-contiguous mapping, reversed & \textbf{0.0820} & $\mathbf{-0.0793}$ & \textbf{lock} \\
\bottomrule
\end{tabular}
\end{table}

\textbf{Best retained revision.} The final kernel, r8, matches the
reference's L1 hit rate and coalescing more closely than the seed does.
Two metrics dominate the residual distance: arithmetic intensity stays
near a quarter of the reference value, and the L2 hit rate reaches
76\% against 91\%.

\subsection{CUDA-First versus Vortex-First Refinement}
\label{sec:evaluation:architecture-sensitivity}

In this section, we explore an alternative design that iterates
on the Vortex implementation directly, instead of refining the CUDA
implementation before translating it to Vortex. In this Vortex-first
design, after each iteration, the Vortex implementation is
translated into a CUDA candidate and profiled on the reference GPU. The
resulting profile differences guide the next Vortex revision.

\begin{figure}[t]
\centering
\begin{subfigure}{\columnwidth}
\centering
\includegraphics[width=\linewidth]{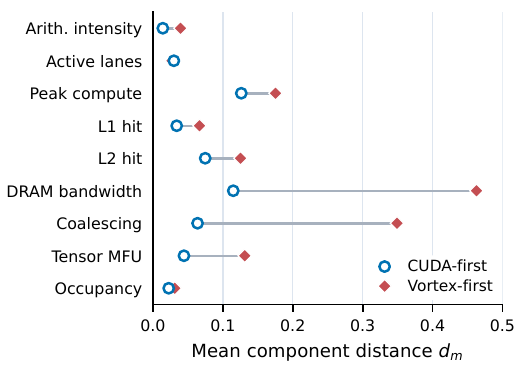}
\caption{Refinement workflow comparison.}
\label{fig:workflow-metric-comparison}
\end{subfigure}\par\medskip
\begin{subfigure}{\columnwidth}
\centering
\includegraphics[width=\linewidth]{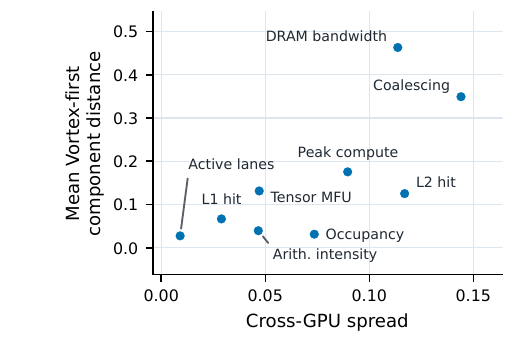}
\caption{Cross-GPU metric variation.}
\label{fig:arch-vs-last-distance}
\end{subfigure}
\caption{Refinement order and architecture-dependent profile differences.}
\label{fig:refinement-order-architecture}
\Description{Panel a compares CUDA-first and Vortex-first mean component distances. Panel b compares cross-GPU metric variation with the same Vortex-first distances. Both panels use all 13 kernels for every plotted metric.}
\end{figure}

Figure~\ref{fig:workflow-metric-comparison} compares the recorded
workflows on H200. For each kernel and workflow, we select
the validated implementation with the lowest aggregate signature distance
and use that implementation for every metric. Each plotted point is the
mean component distance over the same 13 kernels in both workflows.

Compared with the default CUDA-first workflow, Vortex-first
refinement produces higher distances for most metrics. DRAM bandwidth
shows the largest increase (mean distance from 0.115 to 0.463).
Coalescing shows the next-largest increase (0.064 to 0.349). Tensor
utilization, compute utilization, and both cache hit rates also have larger
distances. Active-lane fraction remains nearly unchanged,
while arithmetic intensity and occupancy have small distances under both
workflows.

The higher Vortex-first distances in
Figure~\ref{fig:workflow-metric-comparison} may arise because the CUDA candidate
faithfully preserves Vortex's execution mechanisms and implementation
constraints, even though it runs on an NVIDIA GPU. It therefore carries
choices imposed by Vortex's architecture into the comparison with the
NVIDIA reference. The
measured distance can therefore reflect either a mismatch in workload
behavior or an existing gap introduced by Vortex's architectural
constraints, or both. This makes the feedback less clear about which
implementation changes would better reproduce the reference behavior.
Metrics that depend strongly on architecture and implementation choices
may be more affected by this ambiguity.

We investigate why the distances of some metrics increase
more sharply than others. As shown in Figure~\ref{fig:arch-vs-last-distance},
we carried out an additional experiment profiling the reference kernels
for \benchmarks{} across A100, L40S, H100, and H200. For each metric, the horizontal coordinate averages
normalized distances across GPU pairs within each kernel and then across
kernels. The vertical coordinate uses the same Vortex-first mean component
distance as Figure~\ref{fig:workflow-metric-comparison}. Metrics with
greater cross-GPU variation tend to have larger Vortex-first distances
(Pearson correlation 0.74). DRAM bandwidth and coalescing both have high
variation and large residual gaps, consistent with their large increases
under Vortex-first refinement. Active-lane fraction has little variation
and a small residual gap, consistent with its similar distance under both
workflows.

These observations support refining the CUDA implementation
to bring its measured behavior closer to the reference before translating
it to Vortex. Refinement on the reference platform gives the agent a
clearer signal of behavioral mismatch before target-specific adaptations
are introduced. Reversing the order makes profile matching harder because
the feedback combines behavioral mismatch with the effects of those
adaptations.

\section{Related Work}
\label{sec:related}

\subsection{Compiler-Based Retargeting and Workload Cloning}
\label{sec:related:retargeting}

Source-level retargeting tools translate a program for a different
backend and preserve its semantics by construction. HIPIFY~\cite{hipify} and
SYCLomatic~\cite{syclomatic} rewrite CUDA source for AMD and SYCL targets, CuPBoP~\cite{cupbop, cupbop-amd, softcuda}
compiles CUDA onto processors without a CUDA runtime, and MLIR-based
flows lower a common representation to several backends~\cite{mlir}. Whenever the executed kernel is available as
source and the target exposes comparable primitives, these routes are
preferable to synthesis. Operator-level compilers such as TVM~\cite{tvm}
relax that requirement to an operator graph, but
their autotuners~\cite{tvm-ansor} emit whichever kernel is fastest on the target, so the
result bears no intended relation to the profile of the kernel the
reference stack actually ran. Our setting removes both options: the hot
kernels are vendor library binaries selected at run time, so there is no
source or intermediate representation to retarget, and the goal is a
benchmark whose measured behavior approximates a specific observed
kernel rather than one that merely computes the same function.


The closest prior formulation is workload cloning, which distills a
proprietary application into a distributable benchmark that reproduces its
measured characteristics~\cite{joshi2006cloning,joshi2008distilling}. That
line of work synthesizes whole applications for CPUs from
microarchitecture-independent program characteristics. The same goal has
since been pursued at the granularity of a single GPU kernel, where a
synthetic clone is generated to mimic ``control flow, branch divergence and
memory access behavior''~\cite{yu2015gpgpuminibench} --- the execution
behavior of the kernel, not the values it computes --- and so cannot be run
in place of the original. We keep the goal and change the setting: the unit
is one GPU kernel, the specification is its profile signature plus the
operator it implements, the target is a different ISA, and the artifact must
additionally reproduce the reference output so that it can be substituted
into the original stack and checked.



\subsection{Robotics and ML Benchmark Suites}
\label{sec:related:benchmarks}

Table~\ref{tab:benchmarks} compares the scope of robotics, ML, and
heterogeneous computing benchmark suites. \benchmarks{} focuses on
kernels from profiled robotics workloads, reconstructed as standalone
Vortex benchmarks through output checks and profile-guided refinement.

\begin{table}[!t]
\caption{Scope of benchmark suites.}
\label{tab:benchmarks}
\centering
\footnotesize
\begin{tabularx}{\columnwidth}{@{}>{\raggedright\arraybackslash}p{0.31\columnwidth} >{\raggedright\arraybackslash}X@{}}
\toprule
\textbf{Suite} & \textbf{Benchmark scope} \\
\midrule
MAVBench~\cite{boroujerdian2018mavbench} & End-to-end micro aerial vehicle applications with closed-loop simulation. \\
RobotPerf~\cite{robotperf} & ROS 2 robotics workloads for comparing computing platforms. \\
Rodinia~\cite{che2009rodinia} & Parallel applications and kernels for heterogeneous CPU/GPU systems. \\
MLPerf Inference~\cite{reddi2020mlperf} & Model-level inference performance under defined deployment scenarios. \\
DeepBench~\cite{deepbench} & Basic deep-learning operations, including matrix multiplication and convolution. \\
\textbf{\benchmarks{}} & Robotics kernel benchmarks reconstructed for Vortex with output checks and profile-guided refinement. \\
\bottomrule
\end{tabularx}
\end{table}

\subsection{Robotics Accelerators and Kernel Baselines}
\label{sec:related:accelerators}

Prior architecture works cover important slices of the robotics stack.
Most of them still target one robot type or algorithm family at a time.
Navion~\cite{suleiman2018navion} and
eSLAM~\cite{liu2019eslam} target VIO and ORB-style feature pipelines.
These overlap with the suite's ORB descriptor-extraction and
Hamming-matching kernels. TorchSparse~\cite{torchsparse},
Ada3D~\cite{ada3d}, and PointPillars-style work~\cite{lang2019pointpillars}
overlap the sparse 3D side of the suite. Robomorphic
Computing~\cite{robomorphic}, RoboShape~\cite{roboshape},
cuRobo~\cite{curobo}, and pRRTC~\cite{prrtc} target motion planning
and robot dynamics. Edge-oriented learned inference is covered by
NanoVLA~\cite{nanovla}, Shallow-$\pi$~\cite{shallowpi},
FastDepth~\cite{fastdepth}, and NeuFlow~\cite{neuflow}. These studies show the breadth of robotics kernels, but usually focus on
one robot type or algorithm family at a time.

\subsection{LLM Kernel Generation}
\label{sec:related:kernel-generation}

Benchmarks including KernelBench~\cite{kernelbench},
TritonBench~\cite{tritonbench}, and TritonGym~\cite{tritongym}
evaluate whether LLMs can generate correct and efficient GPU kernels
from PyTorch, Triton, or operator-level specifications. These works are
closest to our use of LLMs as a code-generation mechanism, but their
primary goal is to evaluate generated kernel quality. Our framework instead uses kernels generated by an LLM as
benchmark candidates derived from profiled robotics workloads.
Output equivalence establishes correctness, and profile deviations are
returned to the LLM to synthesize the next candidate.

KForge uses a similar profile-guided LLM refinement loop,
but it starts from available reference implementations and leaves synthesis
for closed-source kernels for future work~\cite{sereda2026kforge}. Kevin and
KernelCraft use iterative feedback as well. Their optimization objective
is to improve execution speed rather than match a reference kernel's
behavior~\cite{baronio2025kevin,nie2026kernelcraft}. CUDA-L2 further
uses reinforcement learning to exceed cuBLAS performance across matrix
multiplication shapes~\cite{su2026cudal2}. This objective differs from
constructing a representative benchmark. Our framework captures key
execution behavior and makes it reproducible on Vortex through a
lightweight process based on correctness checks and profile feedback,
without requiring the original kernel source or additional model training.

\section{Conclusion}
\label{sec:concl}

\framework{} reconstructs portable benchmarks from CUDA workloads
without the original kernel source. It uses workload context,
output-equivalence checks, and profile feedback to generate and refine
kernels for the target architecture.


Using \framework{}, we construct \benchmarks{}, a suite of 26 Vortex/CUDA pairs from
ACT, Diffusion Policy, and Pi0, profiled on H200 and Jetson AGX Orin.
All pairs pass both equivalence checks, and refinement lowers profile
distance in the detailed case studies. 
The suite provides reusable
kernels for Vortex architecture research.
We will release \framework{} and the \benchmarks{} benchmark suite as open source to support benchmark construction on Vortex and other open GPUs.

\begin{acks}
The authors used AI-assisted writing tools to improve
grammar, concision, and clarity.
\end{acks}

\newpage
\bibliographystyle{ACM-Reference-Format}
\bibliography{reference}
\end{document}